\documentstyle[12pt,leqno,bezier]{article}
\def\today{{\rm August 30, 1999}}
\newtheorem{theorem}{Theorem}[section]
\newtheorem{definition}[theorem]{Definition}

\newtheorem{example}[theorem]{Example}

\newtheorem{sublemma}[theorem]{Sublemma}
\newtheorem{proposition}[theorem]{Proposition}
\newtheorem{remark}[theorem]{Remark}
\newtheorem{remarks}[theorem]{Remarks}
\newtheorem{exercise}[theorem]{Exercise}

\catcode`\@=11

\font\tenbold=msbm10 scaled \magstep1
\font\sevenbold=msbm7 scaled \magstep1
\font\fivebold=msbm5 scaled \magstep1
\newfam\boldfam \def\Bbb{\fam\boldfam\tenbold}
\textfont\boldfam=\tenbold
\scriptfont\boldfam=\sevenbold
\scriptscriptfont\boldfam=\fivebold

\def\ps@myheadings{\let\@mkboth\@gobbletwo
\def\@oddhead{\ifnum\count0=1 \hfill\else
\rightmark \hfil \rm\thepage\fi}%
\def\@oddfoot{\ifnum\count0=1 \hfill \rm 1 \hfill \else
\hfill \rm- \today\ - \hfill \fi}
\def\@evenhead%
{\rm\leftmark\hfil\rm\thepage}%
\def\sectionmark##1{}
\def\subsectionmark##1{}}

\def\@begintheorem#1#2{\it \trivlist \item[\hskip
 \labelsep{\bf #1\ #2.}]}
\def\@opargbegintheorem#1#2#3{\it \trivlist\item[\hskip%
 \labelsep{\bf #1\ #2.\ (#3)}]}
\def\@endtheorem{\endtrivlist}

\def\@listI{\leftmargin\leftmargini \parsep 1pt plus 2.5pt
 minus 1pt\topsep 5pt plus 2pt minus 3pt%
 \itemsep 0pt plus 2.5pt minus 1pt}
\let\@listi\@listI
\@listi

\def\@sect#1#2#3#4#5#6[#7]#8{\ifnum #2>\c@secnumdepth%
 \def \@svsec {}\else \refstepcounter {#1}\edef \@svsec%
 {\csname the#1\endcsname. \hskip .1em }\fi \@tempskipa%
 #5\relax \ifdim \@tempskipa >\z@ \begingroup #6\relax%
 \@hangfrom {\hskip #3\relax \@svsec }{\interlinepenalty%
 \@M #8.\par }\endgroup \csname #1mark\endcsname {#7}%
 \addcontentsline {toc}{#1}{\ifnum #2>\c@secnumdepth%
 \else \protect \numberline {\csname the#1\endcsname. }%
 \fi #7}\else \def \@svsechd {#6\hskip #3\@svsec #8.%
 \csname #1mark\endcsname {#7}\addcontentsline {toc}{#1}%
 {\ifnum #2>\c@secnumdepth \else \protect \numberline%
 {\csname the#1\endcsname. }\fi #7}}\fi \@xsect {#5}}

\def\section{\@startsection {section}{1}{\z@ }%
 {-3.5ex plus -1ex minus -.2ex}{2.3ex plus .2ex}{\bf }}

\def\thebibliography#1{%
 \section *{References.\@mkboth {REFERENCES}{REFERENCES}}%
 \list {[\arabic {enumi}]}{\settowidth \labelwidth {[#1]}%
 \leftmargin \labelwidth \advance \leftmargin \labelsep %
 \usecounter {enumi}} \def \newblock %
 {\hskip .11em plus .33em minus -.07em} \sloppy \clubpenalty 4000%
 \widowpenalty 4000 \sfcode`\.=1000\relax}

\def\@maketitle{%
 \newpage \null \vskip 2em
 \begin{center}
{\Large\bf \@title \par }
 \vskip 1.5em
 {\large \lineskip .5em
 \begin {tabular}[t]{c}\@author
 \end{tabular}\par} 
 \end{center}
  \vskip .8em}

\def\abstract{%
\if@twocolumn \section *{Abstract}
 \else \small\quotation\noindent{\bf Abstract.}\fi}

\catcode`\@=13

\def\qed{\hspace*{\fill}
\mbox{\hphantom{mm}\rule{0.25cm}{0.25cm}}\\}

\def\doubless#1#2{{
\def\arraystretch{.5}
\begin{array}{c}
\mbox{\scriptsize $\scriptstyle #1$}
\\
\mbox{\scriptsize $\scriptstyle #2$}
\end{array}\def\arraystretch{1}
}}

\def\Modtriple{{\Bbb M}\hskip .1em} 
\def\twModtriple{{\widetilde {\Modtriple}}}
\def\calh{{\cal H}}

\def\bk{{\bf k}}

\def\D{\Delta}

\def\F{{\cal F}}

\def\calA{{\cal A}}
\def\Leg{{\rm Leg}}
\def\P{{\cal P}}

\def\rr{{\bf r}}
\def\pa{{\partial}}
\def\DD{{\sf D}}
\def\Iso{{\rm Iso \hskip .1em}}
\def\calhrel{{\cal H}_{\rm rel}}
\def\gh{{\rm gh}}
\def\sign#1{{(-1)^{#1}}}
\def\angle#1#2{{\langle#1,#2\rangle}}
\def\mod{{\rm\ mod\ }}
\def\id{1\!\!1}
\def\ext{\mbox{\large$\land$}}
\def\coext{\mbox{${}^c\!\ext$}}
\def\sgn{\mbox{\rm sgn}}
\def\PD#1{\Phi^{#1}_\nabla}
\def\FO#1{\Psi^{#1}_\Omega}
\def\DER#1#2{{\rm Der}^{#1}_{#2}}
\def\coDER#1#2{{\rm coDer}^{#1}_{#2}}
\def\DB{{\overline \Delta}}
\def\coeff{{\rm Coef}}
\def\unsh{{\rm unsh}}
\def\coll#1{\{#1(n)\}_{n\geq 1}}
\def\Hom#1#2{{\rm Hom}(#1,#2)}
\def\Com{\hbox{{$\cal C$}\hskip -.2mm{\it om\/}}}
\def\Ass{\hbox{{$\cal A$}{\it ss\/}}}
\def\Lie{\hbox{{$\cal L$}{\it ie\/}}}

\def\ss{{\bf s\hskip0mm}}

\def\Mod{{\mbox{\rm Mod}}}

\def\suspit#1{{\uparrow^{#1}\,}}
\def\desuspit#1{{\downarrow^{#1}\,}}

\def\Fey{{\sf F}}
\def\fey{{\sf F}}
\def\mod{\mbox{\rm mod }}
\def\znamenko#1{{(-1)^{#1}}}
\def\ot{{\otimes}}
\def\Rada#1#2#3{#1_{#2},\dots,#1_{#3}}
\def\rada#1#2{{#1,\ldots,#2}}
\def\drada#1#2#3{#1_{#2}\cdots#1_{#3}}
\def\orada#1#2#3{#1_{#2}\otimes \cdots \otimes#1_{#3}}
\def\prada#1#2#3{#1_{#2}+\cdots +#1_{#3}}
\def\susp{\uparrow\!}
\def\desusp{\downarrow\!}
\def\c{{\circ}}
\def\frakK{{\frak K}}
\def\({(\hskip -.15em(}
\def\){)\hskip -.15em)}
\def\End{\hbox{${\cal E}\hskip -.1em {\it nd}$}}
\def\otexp#1#2{{#1}^{\ot #2}}
\def\lbigbrace{{\left(\rule{0pt}{12pt}\right.}}
\def\rbigbrace{{\left.\rule{0pt}{12pt}\right)}}
\def\span{{\rm Span}}\def\Span{\span}
\def\bfk{{\bf k}}
\def\MOp{{\tt MOp}}
\def\COp{{\tt COp}}
\def\vert{{\rm Vert}}

\def\ss{{\bf s}}
\def\dz#1#2{{\(#1,#2\)}}
\def\fatGamma{{{\hbox{$\Gamma \hskip-.575em \Gamma \hskip-.575em \Gamma$}}}}
\def\Gammacat#1#2{{\fatGamma}\(#1,#2\)} 
\def\Flag{{\rm Flag}}
\def\Edg{{\rm Edg}}
\def\collc#1{\{#1\(n\)\}_{n\geq 1}}
\def\collm#1{\{#1\(g,n\)\}_{n\geq 1}}

\def\colim#1{\mathop{{\rm colim}}%
             \limits_{\rule{0em}{1em}\mbox{\scriptsize $#1$}}}
\def\isogamma#1#2#3{{\Iso_{{#1}}\Gammacat{#2}{#3}}}
\def\leg{{\rm Leg}}
\def\stackelra#1{{\stackrel{{#1}}{\elra}}}
\def\calC{{\cal C}}
\def\Vert{{\rm Vert}}
\def\elra{{\hbox{$-\hskip-2mm-\hskip-2mm-%
                 \hskip-2mm-\hskip-2mm-\hskip-2mm\longrightarrow$}}}
\def\Det{{\rm Det}}

\def\ModCom{\Mod{(\Com)}}
\def\Orada#1#2#3{#1_{#2}\otimes \cdots \otimes#1_{#3}}
\def\ext{\mbox{\large$\land$}}
\def\Prada#1#2#3{#1_{#2}+\cdots +#1_{#3}}
\def\lbigbrace{{\left(\rule{0pt}{12pt}\right.}}
\def\rbigbrace{{\left.\rule{0pt}{12pt}\right)}}

\font\tengoth=eufm10 scaled \magstep1
\font\sevengoth=eufm7 scaled \magstep1
\font\fivegoth=eufm5 scaled \magstep1
\newfam\gothfam \def\frak{\fam\gothfam\tengoth}
\textfont\gothfam=\tengoth
\scriptfont\gothfam=\sevengoth
\scriptscriptfont\gothfam=\fivegoth

\begin{document}

\pagestyle{myheadings}
\bibliographystyle{plain}
\baselineskip20pt plus 2pt minus 1pt
\parskip3pt plus 1pt minus .5pt

\title{Loop homotopy algebras in closed string field theory}

\author{Martin Markl%
\thanks{This work was supported by the
grant GA \v CR 201/96/0310.}}

\maketitle

\begin{abstract}
Barton Zwiebach constructed~\cite{zwiebach:NuclPh93} 
`string products' on the
Hilbert space of a combined conformal field theory of matter and
ghosts, satisfying the `main identity.' It has been well
known that the `tree level' of the theory gives an example of a
strongly homotopy Lie algebra (though, as we will see later, this is
not the whole truth).

Strongly homotopy Lie algebras are now well-understood objects. On
the one hand, strongly homotopy Lie algebra is given by 
a square zero coderivation
on the cofree cocommutative connected
coalgebra~\cite{lada-stasheff:IJTP93,%
lada-markl:CommAlg95}; on the other
hand, strongly homotopy Lie algebras are algebras 
over the cobar dual of the
operad $\Com$ for commutative algebras~\cite{ginzburg-kapranov:DMJ94}. 

As far as we know, no such characterization of the structure
of string products for arbitrary genera has been available,
though there are two series of papers 
directly pointing towards
the requisite characterization. 

As far as the characterization in terms of (co)derivations is
concerned, we need the concept of {\em higher order (co)derivations\/},
which has been developed, for example, 
in~\cite{akman:preprint,bering-damgaard-alfaro:preprint}. These
higher order derivations were used in the analysis of the `master
identity.' For our characterization we need to understand the behavior
of these higher (co)derivations on (co)free (co)algebras.

The necessary machinery for the operadic approach
is that of {\em modular operads\/}, anticipated
in~\cite{behrend-manin:1995} and introduced 
in~\cite{getzler-kapranov:CompM98}. We believe that 
the modular operad
structure on the compactified moduli space of Riemann surfaces of
arbitrary genera implies the existence of
the structure we are interested in the same manner as was
explained for the tree level in~\cite{kimura-stasheff-voronov:1993}. 

We also indicate how to adapt the loop homotopy structure to
the case of {\em open string field theory\/}~\cite{stasheff3}.

\vskip 2mm
\noindent
{\bf Plan of the paper: }\hskip 1.2mm Section~\ref{introduction} -- 
Introduction
\hfill\break 
\noindent\hphantom{\bf Plan of the paper: } Section~\ref{interlude} -- 
    Sign interlude and the definition \hfill\break 
\noindent\hphantom{\bf Plan of the paper: } Section~\ref{higher} -- 
     Higher order (co)derivations \hfill\break 
\noindent\hphantom{\bf Plan of the paper: } Section~\ref{1st} -- 
Loop homotopy Lie 
     algebras - 1st description \hfill\break 
\noindent\hphantom{\bf Plan of the paper: } Section~\ref{2nd} -- 
Loop homotopy Lie
     algebras - operadic approach \hfill\break 
\noindent\hphantom{\bf Plan of the paper: } 
Section~\ref{generalizations} -- Possible 
     generalizations (open strings)


\vskip 2mm
\noindent
{\bf Classification:} 81Q30, 18C39

\vskip 2mm
\noindent
{\bf Acknowledgment:} I would like to express my gratitude to Jim
Stasheff for reading the manuscript and many helpful remarks and suggestions.
\end{abstract}

\section{Introduction}
\label{introduction}

Let $\calh$ be the Hilbert space of a combined conformal field theory of
matter and ghosts and let $\calhrel \subset \calh$ be the
subspace of elements annihilated by $b_0^- := b_0 -
\overline{b}_0$ and $L_0^- := L_0 - \overline{L}_0$ 
(see, for example,~\cite[Section~4]{kimura-stasheff-voronov:1993}).
Barton Zwiebach constructed in~\cite{zwiebach:NuclPh93}, 
for each `genus' $g\geq 0$
and for each $n\geq 0$, multilinear
`string products' 
\[
\calhrel^{\ot n} \ni \Rada B1n \longmapsto [\Rada B1n]_g \in \calhrel.
\]
Recall the basic properties of these products. If $\gh(-)$ denotes the
ghost number, then~\cite[(4.8)]{zwiebach:NuclPh93}
\[
\gh ([\Rada B1n]_g) = 3-2n + \sum_{i=1}^n \gh(B_i).
\]
The string products are graded (super) 
commutative~\cite[(4.4)]{zwiebach:NuclPh93}:
\begin{equation}
\label{12}
[\Rada B1i,\Rada B{i+1}n]_g = (-1)^{B_i B_{i+1}} 
[\Rada B1{i+1},\Rada B{i}n]_g.
\end{equation}
Here we used the notation
\[
(-1)^{B_i B_{i+1}} := (-1)^{\gh(B_i) \gh(B_{i+1})}.
\]
For $n=0$ and $g\geq 0$, 
$[\, .\, ]_g\in \calhrel$ is just a constant, 
and the products
are constructed in such a way that 
$[\, .\, ]_0 =0$~\cite[(4.6)]{zwiebach:NuclPh93}. The
linear operation $[B]_0 =: QB$ is identified with the BRST
differential of the theory. These product satisfy, for all $n,g$, the
{\em main identity\/}~\cite[(4.13)]{zwiebach:NuclPh93}:
\begin{eqnarray}
\label{0}
\lefteqn{
0=
\sum \sigma(i_l,j_k)\left[\Rada B{i_1}{i_l},[\Rada
B{j_1}{j_k}]_{g_2}\right]_{g_1}}
\\
\nonumber
&& \hskip3cm +
\frac 12 \sum_s (-1)^{\Phi_s} [\Phi_s,\empty \Phi^s,\Rada B1n]_{g-1}. 
\end{eqnarray}
Here the first sum runs over all $g_1+g_2 =g,\ k+l = n$, and all
sequences $i_1 < \cdots <i_l$, $j_1 < \cdots < j_k$ such 
that $\{\Rada
i1l, \Rada j1k\} = \{1,\ldots,n\}$. Such sequences are called
unshuffles (see the terminology introduced at the beginning of
Section~\ref{interlude}). 
The sign  
$\sigma(i_l,j_k)$ is picked up by rearranging the sequence
$(Q,\Rada B1n)$ into the order $(\Rada B{i_1}{i_l},Q,\Rada
B{j_1}{j_k})$.
In the second sum,
$\{\Phi_s\}$ is a basis
of $\calhrel$ and $\{\empty \Phi^s\}$
its dual basis in the sense that
\[
(-1)^{\Phi_r}
\angle {\Phi_r}{\empty \Phi^s} = \delta_r^s
\mbox{ (Kronecker delta),}
\] 
where $\angle --$ denotes the bilinear inner
product on $\calh$~\cite[(2.44)]{zwiebach:NuclPh93}.
Let us remark that, in the original formulation 
of~\cite{zwiebach:NuclPh93}, $\{\Phi_s\}$ was a
basis of the whole $\calh$, but the sum in~(\ref{0}) was 
restricted to $\calhrel$. 
The product satisfies~\cite[(2.62)]{zwiebach:NuclPh93}:
\begin{equation}
\label{1}
\angle AB = \sign{(A+1)(B+1)}\angle BA
\end{equation}
and it is nontrivial only for elements whose ghost numbers 
add up to five:
\begin{equation}
\label{2}
\mbox{if $\angle AB \not = 0$, then $\gh(A)+ \gh(B)=5$.}
\end{equation}
The above two conditions in fact imply that $\angle AB = \angle
BA$. Moreover, the product $\angle --$ is 
$Q$-invariant~\cite[2.63]{zwiebach:NuclPh93}:
\begin{equation}
\label{av}
\angle {QA}B = \sign A \angle A{QB}.
\end{equation}
Conditions~(\ref{1}) and~(\ref{2}) also imply that the element
$\Phi := \sign{\Phi_s}\Phi_s \ot \empty \Phi^s \in \calhrel^{\ot 2}$ 
is symmetric
in the sense that
\begin{equation}
\label{3}
\sign{\Phi_s}\Phi_s \ot \empty \Phi^s = \sign{\Phi_s}\empty 
\Phi^s \ot \Phi_s = - \sign{\empty \Phi^s}\empty \Phi^s \ot \Phi_s.
\end{equation}
We use, in the previous formula as well as at many places in the rest of the
paper, the Einstein convention of summing over
repeated indices.

The last important property of string products is that the element
\begin{equation}
\label{4}
\Phi_s \ot [\empty \Phi^s,\Rada B1{n-1}]_g \in \calhrel^{\ot 2}
\end{equation}
is {\em antisymmetric\/}. This property is not explicitly stated
in~\cite{zwiebach:NuclPh93}, 
though it is used in the proof of the
identity~\cite[(4.28)]{zwiebach:NuclPh93}:
\[
\sum_s
\left[\Rada B1{l},\Phi_s,[\empty 
\Phi^s,\Rada A1{k}]_{g_2}\right]_{g_1} = 0,
\mbox{ for arbitrary $l\geq 0$, $k\geq 0$,}
\]
which then immediately follows from the
antisymmetry~(\ref{4}) by the graded commutativity~(\ref{12})
of string
products. Equation~(\ref{4}) is a consequence of the important fact
that the string products are defined with the aid of the 
{\em multilinear string
functions\/}~\cite[(7.72)]{zwiebach:NuclPh93}
\[
\calhrel^{\ot (n+1)} \ni \Rada B0n \longmapsto \{\Rada B0n\}_g 
\in {\bf C}
\]
by~\cite[(4.33)]{zwiebach:NuclPh93}
\begin{equation}
\label{5}
[\Rada B1n]_g := \sum_t \sign{\Phi_t} \empty \Phi^t \cdot 
\{\Phi_t,\Rada B1n\}_g
\end{equation}
Let us show that 
the graded commutativity~\cite[(4.36)]{zwiebach:NuclPh93}
\[
\{\Rada B0i,\Rada B{i+1}n\}_g = \sign{B_iB_{i+1}}
\{\Rada B0{i+1},\Rada B{i}n\}_g
\]
of the string multilinear
functions implies the antisymmetry of the element in~(\ref{4}).
Indeed, because of~(\ref{3}), we may write~(\ref{5}) as
\[
[\Rada B1n]_g =
\sum_t\sign{\Phi_t}\Phi_t \cdot \{\empty \Phi^t,\Rada B1n\}_g
\]
thus the element in~(\ref{4}) takes the form 
\[
\sum_{s,t} \sign{\Phi_t} (\Phi_s \ot \Phi_t) \cdot \{
\empty \Phi^t,\empty \Phi^s,\Rada B1{n-1}\}_g.
\] 
The antisymmetry we are proving means that
\begin{eqnarray*}
\lefteqn{
\sum_{s,t} \sign{\Phi_t} \Phi_s \ot \Phi_t \cdot \{
\empty \Phi^t,\empty \Phi^s,\Rada B1{n-1}\}_g =\hskip 2cm}
\\
&&
\hskip 2cm -\sum_{s,t} \sign{\Phi_t+ \Phi_s\Phi_t} 
\Phi_t \ot \Phi_s \cdot \{
\empty \Phi^t,\empty \Phi^s,\Rada B1{n-1}\}_g.
\end{eqnarray*}
The replacement $t \longleftrightarrow s$ in the right-hand side of
the above equation gives
\[
-\sum_{s,t} \sign{\Phi_s+ \Phi_t\Phi_s} 
\Phi_s \ot \Phi_t \cdot \{
\empty \Phi^s,\empty \Phi^t,\Rada B1{n-1}\}_g
\]
which can be further rewritten, using the graded commutativity of
string functions, as
\begin{equation}
\label{7}
-\sum_{s,t} \sign{\Phi_s+ \Phi_t\Phi_s+ \empty \Phi^s \empty \Phi^t} 
\Phi_s \ot \Phi_t \cdot \{
\empty \Phi^t,\empty \Phi^s,\Rada B1{n-1}\}_g.
\end{equation}
Since $\gh(\empty \Phi^s) \equiv \gh(\Phi_s)+1\ (\mod 2)$ and 
$\gh(\empty \Phi^t) \equiv \gh(\Phi_t)+1\ (\mod 2)$, 
\[
\gh(\empty \Phi^s)\gh(\empty \Phi^t) \equiv
\gh(\Phi_s)\gh(\Phi_t)+ \gh(\Phi_s)+
\gh(\Phi_t)+1\ (\mod 2),
\]
therefore the sign factor in~(\ref{7}) is $\sign{\Phi_t}$. This
proves the claim.

\section{Sign interlude and the definition}
\label{interlude}

In this brief section we rewrite the axioms of string products into a
more usual and convenient formalism. 
All algebraic objects will be considered over a fixed
field $\bk$ of characteristic zero. 
This, of course, includes the case $\bk = {\Bbb C}$ of the 
previous section.
We will systematically use the
Koszul sign convention meaning that whenever we commute two
`things' of degrees $p$ and $q$, respectively,
we multiply by the sign factor $(-1)^{pq}$. Our conventions concerning
graded vector spaces, permutations, shuffles, etc., will follow
closely those of~\cite{markl:JPAA92}.
For graded indeterminates $\Rada x1n$ and a permutation $\sigma\in
\Sigma_n$ define the {\em Koszul sign\/}
$\epsilon(\sigma)=\epsilon(\sigma;\Rada x1n)$ by
\[
x_1\land\dots\land x_n = \epsilon(\sigma;x_1,\dots,x_n)
\cdot x_{\sigma(1)}\land \dots \land x_{\sigma(n)},
\]
which is to be satisfied in the free graded commutative algebra
$\ext(\Rada x1n)$. Define also 
\[
\chi(\sigma) := \chi(\sigma;\Rada x1n)
:= \sgn(\sigma)\cdot \epsilon(\sigma;\Rada x1n).
\] 
We say that $\sigma
\in \Sigma_n$ is an $(i,j)$-{\em unshuffle\/}, $i+j=n$, if
$\sigma(1)<\cdots<\sigma(i)$ and $\sigma(i+1)< \cdots < \sigma(n)$. 
In this case we write $\sigma \in \unsh(i,j)$.
In the obvious similar manner one may introduce 
$(i,j,k)$-unshuffles, etc.

Let us denote, for a graded vector space $U$, by 
$\susp U$ (resp.~%
$\desusp U$)
the suspension (resp. the desuspension) of $U$, i.e.~the graded
vector space defined by $(\susp U)_p := U_{p-1}$ (resp. $(\desusp U)_p
:= U_{p+1}$).
We have the obvious natural maps $\uparrow : U \to\, \susp U$ and 
$\downarrow : U \to\, \desusp U$.

For a graded vector space $U$, let its {\em reflection\/} $\rr(U)$ 
be the graded vector space defined by $\rr(U)_p := U_{-p}$. There is an
obvious natural map $\rr: U \to \rr(U)$. Observe that $\rr^2 = \id$,
$\rr \hskip.5mm \circ \hskip -.5mm \susp\ =\ \desusp \circ\ \rr$ 
and $\rr  \hskip.5mm \circ\hskip -.5mm  \desusp\ = \susp
\circ\ \rr$.

Take now $V:= \rr(\desusp \calhrel)$. Define, for each $g\geq 0$ and $n\geq
0$, multilinear maps $l^g_n :
V^{\ot n} \to V$ by 
\[
l^g_n(\Rada v1n) := \sign{(n-1)v_1+(n-2)v_2+\cdots + v_{n-1}} \desusp
[\rada{\susp\rr(v_1)}{\susp\rr(v_n)}]_g,
\mbox{ for $\Rada v1n \in V^{\ot n}$}.
\]
Define also the bilinear form $B : V\ot V \to {\Bbb C}$ by
\begin{equation}
\label{las}
B(u,v) :=
\angle {\susp \rr(u)}{\susp \rr(v)}
\end{equation}
and, finally, the element $h = h_s \ot \empty h^s$ by $h_s:=
\sign{\Phi_s}\rr(\desusp \Phi_s)$, $\empty h^s := 
\rr(\desusp \empty \Phi^s)$, which means that
\[
h_s \ot \empty h^s :=\sign{\Phi_s}\rr( \desusp 
\Phi_s) \ot\rr( \desusp \empty \Phi^s)
\mbox{ (Einstein summation convention).}
\]
A technical, but absolutely straightforward, 
calculation shows that the
above structure is an example of a loop homotopy Lie 
algebra in the sense
of the following definition.

\begin{definition}
\label{e}
A {\em loop homotopy Lie algebra\/} 
is a triple $V = (V,B,\{l^g_n\})$
consisting of
\begin{itemize}
\item[(i)]
a ${\Bbb Z}-$graded vector space $V$, $V_* = \bigoplus V_i$,
\item[(ii)]
a graded 
symmetric nondegenerate bilinear degree $+3$ form $B : V \ot V \to
\bk$, and
\item[(iii)]
the set $\{l^g_n\}_{n,g\geq 0}$ of degree $n-2$ multilinear 
antisymmetric operations \mbox {$l^g_n : V^{\ot n}\to V$}.
\end{itemize}
These data are supposed to satisfy the following two axioms.
\begin{itemize}
\item[(A1)]
For any $n,g\geq 0$ and $\Rada v1n \in V$, the 
following `main identity'
\begin{eqnarray}
\label{6}
\label{main-identity}
0\!\!&=& \hskip -6mm
\sum_\doubless{k+l=n+1}{g_1+g_2=g}
\sum_{\sigma \in \unsh(l,n-l)} \hskip -3mm
\chi(\sigma)(-1)^{l(k-1)}l^{g_1}_k(l^{g_2}_l
(\Rada v{\sigma(1)}{\sigma(l)}),\Rada
v{\sigma(l+1)}{\sigma (n)})
\\
\nonumber
&&\hskip 2cm+\frac12 \sum_s 
\sign{h_s+n} l^{g-1}_{n+2}(h_s,\empty h^s,\Rada v1n)
\end{eqnarray}
holds.  In the second term,  
$\{h_s\}$ and $\{\empty h^s\}$ are bases of the vector space
$V$ dual to each other in the sense that
\begin{equation}
\label{14}
B(\empty h^s,h_t) = \delta_{ts}.
\end{equation}
\item[(A2)]
The element
\begin{equation}
\label{acx}
\sign{(n+1)h_s} h_s \ot l^g_n(\empty h^s,\Rada v1{n-1}) \in V\ot V
\end{equation}
is symmetric, for all $g\geq 0,n\geq 0$, and $\Rada v1{n-1}\in V$.
\end{itemize}
\end{definition}

\begin{remarks}
{\rm\
To give a reasonable meaning to the `basis $\{h_s\}$ of $V$,' we must
suppose either that $V$ is finite dimensional, or that it has a
suitable topology, as in the case of string products. We will always
tacitly 
assume that assumptions of this form have been made. 
In the `main identity' for $g=0$ we put, by definition,
$l^{-1}_n=0$. 

Because $\deg(h_s)+\deg(\empty h^s)=-3$, $\deg(h_s)\deg(\empty h^s)$ is
even. The graded symmetry of 
$B$ then implies that, besides of~(\ref{14}), 
also $B(h_s, \empty h^t)=
\delta_{st}$. The element $h = h_s\ot \empty h^s$ is easily seen to be
symmetric, $ h_s\ot \empty h^s =  \sign{h_s\empty h^s}\empty h^s\ot
h_s = \empty h^s\ot
h_s$.

For $n=0$ axiom~(\ref{0}) gives
\[
0= 
\sum_{g_1+g_2 =g} l^{g_1}_1(l^{g_2}_0(.)) + \frac12 \sum_s 
\znamenko{h_s} l^{g-1}_2(h_s,h^s),
\]
while for $n=1$ it gives
\begin{equation}
\label{cameras}
0 = \sum_{g_1+g_2 =g}
(l^{g_1}_1(l^{g_2}_1(v))+ l^{g_1}_2(l^{g_2}_0(.),v))
- \frac 12\sum_s\znamenko{h_s}l_3^{g-1}(h_s,h^s,v),\ v \in  V.
\end{equation}
{}From this moment on, we will assume that $l^g_0 =0$, for all $g\geq
0$, that is, the theory has `no constants.' This assumption is not
really necessary, but it will  considerably simplify our exposition.
}\end{remarks}

\begin{exercise}
{\rm
Let us denote $\pa:= l^0_1$. Equation~(\ref{cameras}) implies
that $\pa^2=0$ (recall our assumption $l^g_0 =0$!). 
Thus $\pa$ is a degree
$-1$
differential
on the space $V$. The symmetry of $h_s \ot \pa(\empty h^s)$
(axiom~(A2) with $n=1$ and $g=1$) is equivalent to the
$d$-invariance of the form 
$B$, $B(\pa u,v)+\sign{u}B(u,\pa v)=0$, for $u,v\in V$.
}\end{exercise}

\noindent 
{\bf The tree level.}
Let us discuss the `tree level' ($g=0$) specialization of the above
structure. The only nontrivial $l^g_n$'s are $l_n := l_n^0$, $n\geq
1$.
The main identity~(\ref{main-identity}) for $g=0$ reduces to
\begin{equation}
\label{15}
0=
\sum_{k+l=n+1}\sum_{\sigma\in \unsh(l,n-l)}
\chi(\sigma)(-1)^{l(k-1)}l_k(l_l
(\Rada v{\sigma(1)}{\sigma(l)}),\Rada
v{\sigma(l+1)}{\sigma (n)})
\end{equation}
while, for $g=1$ it
gives (after forgetting the overall factor $\frac{\sign n}2$)
\begin{equation}
\label{115}
0= \sum_s 
\sign{h_s} l_{n+2}(h_s,\empty h^s,\Rada v1n).
\end{equation}
Axiom~(A2) says that the elements
\begin{equation}
\label{17}
\sign{(n+1)h_s} h_s \ot l_n(\empty h^s,\Rada v1n)
\end{equation}
are symmetric. We immediately recognize~(\ref{15}) as 
the defining equation
for {\em strongly homotopy Lie algebras\/} 
\cite[Definition~2.1]{lada-markl:CommAlg95}. 
Thus the tree level
loop homotopy Lie algebra is a strongly 
homotopy Lie algebra $(V,\{l_n\})$ with
an additional structure given by a bilinear form $B$ such that
the element $h = h_s\ot \empty h^s$, uniquely determined by $B$,
satisfies~(\ref{115})
and~(\ref{17}). We see that the `tree-level' 
specialization is a richer
structure than just a strongly homotopy Lie algebra as 
it is usually
understood. 
A proper name for such a structure would be
a {\em cyclic\/} strongly
homotopy Lie algebra.

\section{Higher order (co)derivations}
\label{higher}

In this section we investigate properties of higher order
coderivations of cofree cocommutative coalgebras. Because this paper is
meant for humans, not for robots, we derive necessary properties
for {\em derivations\/} on free commutative algebras, and then simply
dualize the results. This is an absolutely correct procedure, except
one fine point related to the cofreeness, see Remark~\ref{Lisegacev}.
The following definitions were taken 
from~\cite{akman:preprint97,bering-damgaard-alfaro:preprint}.

Let $A$ be a graded (super) commutative algebra and $\nabla: A \to A$ a
homogeneous degree $k$ linear map.
We define inductively, for each $n\geq 1$, degree $k$ 
linear {\em deviations\/}
$\PD n: A^{\ot n}\to A$ by
\begin{eqnarray*}
\PD1(a) &:=& \nabla(a),
\\
\PD2(a,b)&:=& \nabla(ab) - \nabla(a)b - \sign {ka} a\nabla(b),
\\
\PD3(a,b,c) &:=&
\nabla(abc) - \nabla(ab)c - \sign{a(b+c)}\nabla(bc)a -
\sign{c(a+b)}\nabla(ca)b 
\\
&&+
\nabla(a)bc + \sign{a(b+c)}\nabla(b)ca + \sign{c(a+b)}\nabla(c)ab,
\\
&\vdots&
\\
\PD{n+1}(\Rada a1{n+1})&:=&
\PD n(a_1,\ldots,a_na_{n+1})- \PD n(\Rada a1n)a_{n+1}
\\
&&-\sign{a_n\cdot a_{n+1}}\PD n(\Rada
a1{n-1},a_{n+1})a_n.  
\end{eqnarray*}
As a matter of fact, it is possible to give a non-inductive formula
for $\PD n$, namely
\begin{equation}
\label{Siso}
\PD n(\Rada a1n) = 
\sum_\doubless{1\leq i \leq n}{\sigma \in \unsh(i,n-i)}
\sign{n-i}\epsilon(\sigma)\nabla(\drada x{\sigma(1)}{\sigma(i)})
\drada x{\sigma(i+1)}{\sigma(n)}.
\end{equation}
We say that $\nabla$ is a {\em derivation of order $r$\/} if $\PD
{r+1}$ is identically zero. In this case we write $\nabla \in \DER
rk(A)$, where $k = \deg(\nabla)$. In the following proposition, which
was stated in~\cite{akman:preprint97}, $[-,-]$ denotes the graded
anticommutator of endomorphisms.

\begin{proposition}
\label{b}
The subspaces $\DER rk(A)$ satisfy:
\begin{itemize}
\item[(i)]
$\DER 1k(A) \subset \DER 2k(A) \subset \DER 3k(A) \subset \cdots$
\item[(ii)]
$\DER rk (A)\circ \DER sl (A)\subset \DER {r+s}{k+l}(A)$, and
\item[(iii)]
$[\DER rk(A),\DER sl(A)] \subset \DER {r+s-1}{k+l}(A)$.
\end{itemize}
\end{proposition}

Let now $A = \ext X$ be the free graded commutative algebra on the
graded vector space $X$. Let us prove the following useful proposition.

\begin{proposition}
\label{a}
Let $\nabla \in \DER rk(\ext X)$. Then $\nabla$ is uniquely determined
by its values on the products
$x_1\cdots x_s$, $s\leq r$, $x_i \in X$ for $1\leq i
\leq s$. In particular, 
\[
\mbox{$\nabla = 0$ if and only if $\nabla(x_1\cdots x_s)=0$, for 
$x_1\cdots x_s$ as above.}
\]
\end{proposition} 

\noindent 
{\bf Proof.}
Since $\nabla \in \DER rk(\ext X)$ is linear, it is enough to prove that
$\nabla(\drada x1s)= 0$ for all $s\leq r$ 
implies that $\nabla(\drada x1n)=0$ for each $n$.
This we prove inductively. Suppose we already know
$\nabla(\drada x1k)=0$, for each $k\leq n$, $n\geq r$, and consider
$\nabla (\drada x1{n+1})$. We compute from~(\ref{Siso}) that
\begin{eqnarray*}
\lefteqn{
\PD{n+1}(\Rada x1{n+1}) =} 
\\
&=&\nabla(\drada x1{n+1}) + 
\hskip -8mm
\sum_\doubless{1\leq i \leq n}{\sigma \in \unsh(i,n-i+1)} 
\hskip -6mm
\sign{n-i+1}\epsilon(\sigma) \nabla(\drada
x{\sigma(1)}{\sigma(i)}) \drada x{\sigma(i+1)}{\sigma(n+1)},
\end{eqnarray*}
Since $\nabla \in \DER rk(\ext X)$ and $n\geq r$,
$\PD{n+1}(\Rada x1{n+1})=0$, 
while the terms in the sum are zero by the inductive
assumption. Thus $\nabla(\drada x1{n+1})=0$ and 
the induction may go on.%
\qed

\begin{remark}
{\rm\
$1$-derivations are ordinary derivations, $\DER 1k(A)
=\DER{}k(A)$. Proposition~\ref{a} then states the standard fact 
that derivations on
free algebras are given by their restrictions to the space of
generators.
}\end{remark}

For a
fixed $n$, we denote by $\ext^n X$ the subspace of $\ext X$ spanned by
the products $\drada x1n$, $x_i \in X, 1 \leq i\leq n$; 
we put, by definition, $\ext^0X:=
\bk$. Let $\iota_n : \ext^n X \hookrightarrow \ext X$ be the
inclusion. The following proposition says that $r$-derivations of
the free algebra $\ext X$ are in one-to-one correspondence with
$r$-tuples of linear maps, $\{f_s : \ext^s X \to \ext X\}_{1\leq s \leq
r}$.

\begin{proposition}
\label{c}
Suppose we are given homogeneous degree $k$ linear maps $f_s :
\ext^s X\to \ext X$, for $1\leq s\leq r$. Then there exists a unique
order $r$ derivation $\nabla \in \DER rk(\ext X)$ such that
\begin{equation}
\label{842}
\nabla \circ \iota_s = f_s, \mbox{ for $1\leq s\leq r$}.
\end{equation}
\end{proposition} 

\noindent 
{\bf Proof.}
The uniqueness follows immediately from Proposition~\ref{a}. To prove
the existence, observe first that, given degree $k$ linear maps $g_s:
\ext^s X\to \ext X$, $1\leq s\leq r$, the formula
\[
\nabla(\drada x1n) :=
\sum_{
\doubless{1\leq s \leq {\rm min}(r,n)}
{\sigma \in \unsh(s,n-s)}
} \epsilon(\sigma)g_s(\drada
x{\sigma(1)}{\sigma(s)}) \drada x{\sigma(s+1)}{\sigma(n)},
\]
defines 
an order $k$ derivation.
Condition~(\ref{842}) then leads to the following system of equations:
\begin{eqnarray*}
f_1(x_1) &=& g_1(x_1),
\\
f_2(x_1x_2) &=& g_2(x_1x_2) + g_1(x_1)x_2 +
\sign{x_1x_2}g_1(x_2)x_1,
\\
&\vdots&
\\
f_r(\drada x1r) &=& \sum_{
\doubless{1\leq s \leq r}
{\sigma \in \unsh(s,n-s)}}
\epsilon(\sigma)g_s(\drada
x{\sigma(1)}{\sigma(s)}) \drada x{\sigma(s+1)}{\sigma(r)}.
\end{eqnarray*}
This system can obviously be solved for $g_s$,  $1\leq s\leq r$.%
\qed

Let us turn our attention to coalgebras. Suppose that $C = (C,\Delta)$
is a cocommutative coassociative coalgebra. To define higher-order
coderivations of $C$, we need analogs of the 
deviations $\PD r$ introduced
above. By duality, we define, for any homogeneous degree $k$
linear
endomorphism $\Omega$ of $C$, 
degree $k$ multilinear maps $\FO n
: C \to C^{\ot n}$ inductively as
\begin{eqnarray*}
\FO 1 &:=& \Omega,
\\
\FO 2 &:=& \Delta \circ\Omega - (\Omega \ot \id)\circ \Delta - 
(\id \ot
\Omega) \circ \Delta,
\\
\FO 3 &:=&
\Delta^{[3]} \c \Omega
-(\D \ot \id) \c (\Omega \ot \id)\c\D - 
T_{312}\c(\D\ot\id) \c(\Omega \ot \id)\c\D
-T_{231}\c(\D\ot\id)\c(\Omega \ot \id)\c\D
\\
&&
+(\Omega \ot \id^2)\c \D^{[3]}+
T_{312}\c (\Omega \ot \id^2)\c \D^{[3]}
+T_{231}\c (\Omega \ot \id^2)\c \D^{[3]}
\\
&\vdots&
\\
\FO{n+1}&:=& (\id^{n-1}\ot \Delta)\circ\FO n
- (\FO n\ot \id) \circ \Delta -
T_{1,2,\ldots,n-1,n+1,n}\circ (\FO n \ot \id)\circ \Delta,
\end{eqnarray*}
where 
\[
\D^{[3]}:= (\D\ot \id)\D\ 
(= (\id \ot \D)\D\mbox { by the coassociativity})
\]
and, for $\sigma \in \Sigma_n$, $T_{\sigma(1)\cdots\sigma(n)} :
C^{\ot n} \to C^{\ot n}$ is defined by
\[
T_{{\sigma(1)}\cdots{\sigma(n)}} (\orada x1n) :=
\epsilon(\sigma)
(\orada x{\sigma(1)}{\sigma(n)}. 
\]
We say that a linear map $\Omega :C \to C$ is an {\em order $r$
coderivation\/}, if $\FO{r+1}$ is identically zero. 
Let $\coDER rk(C)$ be
the space of all such maps. The following proposition is an exact dual
of Proposition~\ref{b}.

\begin{proposition}
\label{f}
The subspaces $\coDER rk(C)$ satisfy:
\begin{itemize}
\item[(i)]
$\coDER 1k(C) \subset \coDER 2k(C) \subset \coDER 3k(C) \subset \cdots$
\item[(ii)]
$\coDER rk (C)\circ \coDER sl (C)\subset \coDER {r+s}{k+l}(C)$, and
\item[(iii)]
$[\coDER rk(C),\coDER sl(C)] \subset \coDER {r+s-1}{k+l}(C)$.
\end{itemize}
\end{proposition}

Let $W$ be a graded vector space and consider again the free graded
commutative algebra $\ext W$ on $W$. We introduce on $\ext W$ a
cocommutative coassociative comultiplication $\D = \id \ot 1 + \DB +
1\ot \id$ by defining the reduced diagonal $\DB$ as
\[
\DB(w_1\cdots w_n)=\sum_{1\leq i\leq n-1}\sum_\sigma
\epsilon(\sigma)(w_{\sigma(1)}\cdots w_{\sigma(i)})\otimes
(w_{\sigma(i+1)}\cdots w_{\sigma(n)}),\
\drada w1n \in \ext^nW,
\]
where $\sigma$ runs through all $(i,n-i)$ unshuffles. We denote the
coalgebra $(\ext W,\D)$ by $\coext W$.

\begin{remark}
\label{Lisegacev}
{\rm
Here it must be pointed out that $\coext W$ is not the cofree
cocommutative coassociative coalgebra cogenerated by $W$, as it is
generally supposed to be. It is the cofree coalgebra in the category of
{\em connected\/} coalgebras, see the discussion 
in~\cite[page~2150]{lada-markl:CommAlg95}.
}
\end{remark}

Denote by $\pi_n : \coext W \to \ext^n W$ the natural projection of
vector spaces. The
following theorem is the exact dual of Proposition~\ref{c}.

\begin{proposition}
\label{d}
For each $r$-tuple $u_s :
\coext W\to \ext^s W$, $1\leq s\leq r$,
of homogeneous degree $k$ linear maps there exists a unique
order $r$ coderivation $\Omega \in \coDER rk(\coext W)$ such that
\begin{equation}
\pi_s \c \Omega = u_s, \mbox{ for $1\leq s\leq r$}.
\end{equation}
\end{proposition}

\section{Loop homotopy Lie algebras - 1st description}
\label{1st}

We already observed at the end of Section~\ref{interlude}
that strongly homotopy Lie algebras are
closely related to 
the `tree level' specializations of loop homotopy Lie
algebras. Recall~\cite[Theorem~2.3]{lada-markl:CommAlg95} 
that strongly homotopy Lie algebras have the following
characterization. 

\begin{proposition}
\label{wv}
There exists a one-to-one correspondence between strongly homotopy
Lie algebra structures on a graded vector space $V$ and degree $-1$
coderivations $\delta \in \coDER{}{-1}(\coext W)$, $W := \susp
V$, with the property $\delta^2=0$.
\end{proposition}

In this section we give a similar characterization for loop homotopy
Lie algebras. Suppose that the vector space $V$ and the bilinear form
$B$ is the same as in Definition~\ref{e}. Let $h = h_s\ot \empty h^s
\in (V\ot V)_{-3}$ be as in~(\ref{14}) (of course, 
$h$ is uniquely determined
by the nondegenerate form $B$).

Let $W := \susp V$ and $y = y_s\ot \empty y^s 
:= 
\susp h_s \ot \susp h^s \in (W\ot W)_{-1}$. Because 
$h$ is symmetric, $y$ is symmetric as well, thus, 
in fact, $y = y_s\empty y^s 
\in (\ext^2 W)_{-1}$. Let us
consider the extension $\coext
W[t]$ of $\coext W$ over the polynomial ring $\bk[t]$, 
$\coext W[t] := \coext W \ot_{\bk} \bk [t]$. By Proposition~\ref{d},
there exist a 
unique coderivation $\theta \in \coDER2{-1}(\coext W[t])$
such that 
\begin{equation}
\label{ee}
\pi_1 ( \theta) = 0
\mbox{ and } 
\pi_2(\theta)(w)=
\left\{
\begin{array}{ll}
0, & w\in \ext^n W[t],\ n > 0,
\\
\frac 12 ty, & w=1 \in \ext^0 W \cdot t^0 \cong \bk.
\end{array}
\right.
\end{equation}
The r\^ole of $\theta$ is to incorporate the form $B$ into our theory.
In the rest of this section we prove the following theorem.

\begin{theorem}
\label{ff}
Under the above notation, there is a one-to-one correspondence
between loop homotopy Lie algebra structures on the graded vector
space $V$ and degree $-1$
coderivations $\delta \in \coDER 1{-1} (\coext W[t])$ such that
\begin{equation}
\label{711}
(\delta + \theta)^2=0.
\end{equation}
\end{theorem}

\noindent 
Let us analyze equation~(\ref{711}). It is, of course, equivalent to
\begin{equation}
\label{ctverec}
\delta^2 + \theta  \delta + \delta \theta + \theta^2 =0.
\end{equation}

\begin{sublemma}
\label{qq}
Under the above notation,
$\theta^2 = 0$, $\delta^2 \in \coDER1{-2} (\coext W[t])$, and
$(\theta  \delta + \delta \theta)\in \coDER2{-2} (\coext W[t])$.
\end{sublemma}

\noindent 
{\bf Proof.}
For $\drada w1n \in \ext^n W$ obviously
\begin{equation}
\label{10}
\theta(\drada w1n) =\frac12 t y_s\empty y^s \drada w1n,
\end{equation}
thus
\begin{equation}
\label{8}
\theta^2(\drada w1n) = \frac14
t^2 y_s\empty y^s y_t \empty y^t \drada w1n. 
\end{equation}
The graded commutativity implies that 
\[
y_s\empty y^s y_t \empty y^t =
\sign{(y_s + \empty y^s)(y_t + \empty y^t)}y_t\empty y^t y_s \empty
y^s = -y_t\empty y^t y_s \empty
y^s.
\] 
On the other hand, the substitution $s \leftrightarrow t$
gives $y_s\empty y^s y_t \empty y^t = y_t\empty y^t y_s \empty
y^s$, therefore $y_t\empty y^t y_s \empty
y^s = 0$, and $\theta^2=0$ by~(\ref{8}).

The remaining two statements follow from Proposition~\ref{f}(iii) and
the observation that
$\delta^2 = \frac 12 [\delta,\delta]$ and $\theta \delta + 
\delta \theta = [\delta,\theta]$.%
\qed 

\noindent
By Sublemma~\ref{qq}, (\ref{ctverec}) reduces to
\begin{equation}
\label{16}
\delta^2 + \theta\delta + \delta\theta = 0.
\end{equation}
By the same sublemma and Proposition~\ref{b}(i), 
$\delta^2 + \theta\delta + \delta\theta$ is
an order $2$ coderivation. Thus~(\ref{16}) is, by Proposition~\ref{d},
equivalent to
\begin{eqnarray}
\label{1u}
\pi_1(\delta^2 + \theta \delta + \delta \theta) &=& 0, 
\mbox{ and}
\\
\label{2u}
\pi_2(\delta^2 + \theta \delta + \delta \theta) &=& 0.
\end{eqnarray} 
Because, by~(\ref{ee}), $\pi_1(\theta) = 0$, equation~(\ref{1u}) further
reduces to 
\begin{equation}
\label{1uu}
\pi_1(\delta^2 +\delta \theta)=0.
\end{equation}
To understand better the meaning of this equation, let us introduce, for
any $g\geq 0$ and $n \geq 0$, linear maps $\delta^g_n : \ext^n W \to
W$ by
\begin{equation}
\label{aa}
\delta^g_n (\drada w1n) := \coeff_g (\pi_1 \delta(\drada w1n)),\
\drada w1n \in \ext^nW,
\end{equation}
where $\coeff_g (-)$ is the coefficient at
$t^g$. By Proposition~\ref{d}, the set $\{\delta^g_n\}_{n,g\geq 0}$
uniquely 
determines the coderivation $\delta$. The explicit formula is 
(compare explicit formulas for coderivations acting on coalgebras 
in~\cite{lada-stasheff:IJTP93}):
\begin{equation}
\label{9}
\delta(\drada w1n)= \sum_{0\leq i \leq n}\epsilon(\sigma)
t^g\delta^g_i (\drada w{\sigma(1)}{\sigma(i)})\drada 
w{\sigma(i+1)}{\sigma(n)},
\end{equation}
where the summation is taken over all $g\geq 0$ and all
$\sigma \in \unsh(i,n-i)$. From this 
and~(\ref{10}) we obtain 
\begin{eqnarray}
\label{11}
\lefteqn{\pi_1 (\delta^2 +\delta\theta)(\drada w1n) = \hskip 2cm}
\\
\nonumber
&&\hskip 1.5cm=
\sum_\doubless{k+l=n+1}{g_1+g_2= g} 
\sum_{\sigma \in \unsh(l,n-1)}\epsilon(\sigma)
t^g \delta^{g_1}_k(\delta^{g_2}_l 
(\drada w{\sigma(1)}{\sigma(l)})\drada 
w{\sigma(l+1)}{\sigma(n)}) 
\\
\nonumber
&&\hskip 2cm +
\frac 12 \sum_{s,g\geq 0} t^{g+1} \delta^{g}_{n+2}
(y_s,\empty y^s,\Rada w1n).
\end{eqnarray}
We formulate the result as:

\begin{sublemma}
Equation~(\ref{1uu}) means that, for all $n\geq 0$, 
$\drada w1n\in \ext^nW$ and
$g\geq 0$,
\begin{eqnarray}
\label{1uuu}
\hskip 4mm 0&=&
\hskip -5mm
\sum_\doubless{k+l=n+1}{g_1+g_2= g} 
\sum_{\sigma \in \unsh(l,n-1)}\hskip -5mm
\epsilon(\sigma)
\delta^{g_1}_k(\delta^{g_2}_l (\drada w{\sigma(1)}{\sigma(l)})\drada 
w{\sigma(l+1)}{\sigma(n)}) \hskip 2.5cm
\\
\nonumber 
&&\hskip 5mm
+\frac 12 \sum_s  \delta^{g-1}_{n+2}(y_s,\empty y^s,\Rada w1n).
\end{eqnarray}
\end{sublemma}
We will see that equation~(\ref{1uuu}) 
will correspond to the `main identity'~(\ref{6}).
Let us make a similar analysis of
equation~(\ref{2u}). Because clearly $\pi_2 (\theta
\delta) = 0$, it reduces to 
\begin{equation}
\label{2uu}
\pi_2(\delta^2 + \delta \theta) = 0.
\end{equation}
Using the similar arguments as above, we obtain, for any $g\geq 0$
and $\drada w1n \in \ext^n W$,
\begin{eqnarray}
\label{x}
\lefteqn{\coeff_g(\pi_2 (\delta^2)(\drada w1n)) =}
\\
\nonumber 
&&=\hskip -5mm
\sum_\doubless{k+l=n+1}{g_1+g_1=g} 
\sum_{\sigma \in \unsh(l,n-l-1,1)}
\hskip -5mm
\epsilon(\sigma)
\delta^{g_1}_k(\delta^{g_2}_l (\drada w{\sigma(1)}{\sigma(l)})\drada 
w{\sigma(l+1)}{\sigma(n-1)})w_{\sigma(n)}
\\
\nonumber 
&&
\hskip 2mm
+
\hskip -5mm\sum_\doubless{p+g=n}{g_1+g_1=g}
\hskip -1mm
\sum_{\sigma \in \unsh(p,q)}
\hskip -5mm
\sign{\prada w{\sigma(1)}{\sigma(p)}} \epsilon(\sigma)
\delta^{g_1}_p(\drada w{\sigma(1)}{\sigma(p)})
\delta^{g_2}_q(\drada w{\sigma(p+1)}{\sigma(n)})
\end{eqnarray}
Similarly, we have
\begin{eqnarray}
\label{y}
\lefteqn{\coeff_g(\pi_2 (\delta \theta)(\drada w1n)) =}
\\
\nonumber
&&
=\frac12\sum_{1\leq i \leq n}
\sign{w_i(\prada w{i+1}n)}
\delta^{g-1}_{n+1}(y_s\empty y^s\drada w1{i-1}\drada
w{i+1}n)w_i 
\\
&&
\hskip 1cm+\nonumber  
\frac12\sum_s
\sign{y_s(\empty y^s+
\prada w1n)} \delta^{g-1}_{n+1}(\empty y^s\drada w1n)y_s
\\
&&
\nonumber 
\hskip 1cm+\frac12\sum_s
\sign{\empty y^s(\prada w1n)} 
\delta^{g-1}_{n+1}(y_s\drada w1n)\empty y^s.
\end{eqnarray}
Now, assuming~(\ref{1uuu}), 
it is immediate to see that the first
term at the right hand side of~(\ref{x}) is minus the first term
at the right hand side of~(\ref{y}). The symmetry $y_s \empty y^s=
\sign{y_s \empty y^s} \empty y^s y_s$ implies that
the second and third terms at the left hand side of~(\ref{y}) are the
same, both equal to $ 1/2 \sum_s\sign{y_s}y_s 
\delta^{g-1}_{n+1}(\empty
y^s\drada w1n)$.
We formulate these observations as

\begin{sublemma}
Assuming~(\ref{1uuu}), equation~(\ref{2uu}) is equivalent to
\begin{equation}
\label{2uuuuu}
\frac 12 \sum_s\sign{y_s}y_s \delta^{g-1}_{n+1}(\empty
y^s\drada w1n)=0, 
\end{equation}
Since we work in the free commutative algebra, ~(\ref{2uuuuu})
is equivalent to the antisymmetry of
\begin{equation}
\label{2uuuuuu}
\frac 12 \sum_s\sign{y_s}y_s \ot \delta^{g-1}_{n+1}(\empty
y^s\drada w1n) \in W\ot W.
\end{equation}
\end{sublemma}

\noindent 
{\bf Proof of Theorem~\ref{ff}.}
Recall that $W = \susp V$. The correspondence between the structure
operations $\{l^g_n\}_{g,n\geq 0}$ 
of a loop homotopy Lie algebra and
coderivations $\delta$ of Theorem~\ref{ff} is given by
\[
l^g_n(\Rada v1n)=
\sign{(n-1)v_1+\cdots +v_{n-1}}
\desusp \delta^g_n(\drada{\susp v}1n),\
\Rada v1n\in V,
\] 
with the inverse formula
\[
\delta^g_n(\drada w1n) =
\sign{{n(n-1)}/2}\sign{(n-1)w_1+\cdots+ w_{n-1}}
\susp l^g_n(\Rada{\desusp w}1n),\ 
\drada w1n \in \ext^n W.
\]
where the multilinear maps $\{\delta^g_n\}$ were introduced in~(\ref{aa}).
Observe the sign $\sign{{n(n-1)}/2}$ in the second formula; it is 
typical for formulas of this type,
see~\cite[Example 1.6]{markl:JPAA92}. 
A routine calculation shows that the substitution
$l^g_n \leftrightarrow \delta^g_n$ converts~(\ref{1uuu}) to~(\ref{6})
and that the symmetry of the element in~(\ref{acx}) is equivalent
to the antisymmetry of the element of~(\ref{2uuuuuu}).%
\qed

\section{Loop homotopy Lie algebras - operadic approach}
\label{2nd}

In this section we give an operadic characterization of loop homotopy
Lie algebras.

We will not repeat here all details of necessary 
definitions concerning operads,
because it would stretch the 
paper beyond any reasonable
limit.
Operads are introduced in the 
classical book~\cite{may:1972}. 
The (co)bar construction over a (co)operad is
defined in~\cite{ginzburg-kapranov:DMJ94}, 
see also~\cite{getzler-jones:preprint}. Cyclic operads
are introduced in~\cite{getzler-kapranov:CPLNGT95} 
while modular operads and the corresponding
modular (co)bar construction (called the Feynman transform) 
in~\cite{getzler-kapranov:CompM98}.
There is also a nice overview~\cite{kapranov:DM}.
These sources are easily available, we will thus rely on them and
indicate only basic ideas.

Recall that a {\em collection\/} is a system
$E = \coll E$ of graded vector spaces such that each $E(n)$ possesses a
right action of the symmetric group $\Sigma_n$.
Any collection $E$ extends  to a functor 
(denoted by the same symbol) from
the category of finite sets to the category of graded vector spaces
with the property that
$E(n) = E(\{\rada 1n\})$~\cite[1.3]{getzler-jones:preprint}.

Let ${\tt T}^{\rm r}_n$ denote the set of rooted (= directed) trees with
$n$ labelled leaves. For a tree $T \in {\tt T}^{\rm r}_n$ and a collection
$E$, denote~(\cite[1.2.13]{ginzburg-kapranov:DMJ94})
\[
E(T) := \bigotimes_{v \in \vert(T)}E({\rm In}(v)),
\] 
where
$\vert(T)$ is the set of the vertices of $T$ and ${\rm In}(v)$ the
set of incoming edges of $v$. 
The {\em free operad\/} on 
$E$~\cite[2.1.1]{ginzburg-kapranov:DMJ94} is then the
collection
\[
\F(E)(n) := \bigotimes_{T\in {\tt T}^{\rm r}_n} E(T),\ n \geq 1,  
\]
with the operadic structure 
induced by the grafting of underlying trees.

Let $\P$ be an operad. Consider the free operad $\F(\desusp \ss
\P^*)$ on the collection 
\[
\desusp \ss
\P^*(n) := \hskip 1mm \susp^{n-2}\P^*(n),\ n\geq 1,
\]
where ${(-)}^*$ is the linear dual.
As proved in~\cite[3.2]{ginzburg-kapranov:DMJ94}, structure
operations of the operad $\P$ induce a differential
$\pa_\DD$ on $\F(\desusp \ss \P^*)$. The differential operad 
$\DD(\P): = (\F(\desusp \ss
\P^*),\pa_\DD)$ is called the (operadic) {\em cobar dual\/}
of the operad $\P$.
It is well-known~\cite[4.2.14]{ginzburg-kapranov:DMJ94} 
that `classical' strongly homotopy Lie algebras are
characterized as follows.

\begin{proposition}
\label{vw}
Strongly homotopy Lie algebras are algebras over the cobar dual
$\DD(\Com)$ of the operad $\Com$ for commutative algebras.
\end{proposition}

The above proposition means that a strongly homotopy Lie algebra
structure on a differential graded vector space $V = (V,\pa)$ is the
same as a morphism $a : \DD(\Com) \to \End_V$ from the operad
$\DD(\Com)$ to  the
endomorphism operad $\End_V$ of $V$~\cite[1.2.9]{ginzburg-kapranov:DMJ94}.

Our aim is to give a similar characterization of loop
homotopy Lie algebras, based on a certain
generalization of operads, called {\em modular} operads.

An intermediate step between ordinary operads and 
modular operads are cyclic operads whose
definition we briefly recall. A {\em cyclic collection\/} is a system
$E =\collc E$ of graded vector spaces such that each $E \(n\)$ has a
right $\Sigma_{n+1}$-action. Each cyclic
collection $E$ induces a functor from the category of finite sets into
the category of
graded vector spaces (denoted again by $E$) such that $E\(\{\rada
0n\}\) = E\(n\)$. This notation differs from
that of~\cite{getzler-kapranov:CPLNGT95} 
and~\cite{behrend-manin:1995} 
where $E\(\{\rada 0n\}\) =E\(n+1\)$.

Let ${\tt T}^{\rm ur}_n$ denote the set of {\em unrooted\/} 
trees $T$ with leaves indexed by
$\{\rada 0n\}$. For a cyclic collection $E$ and a tree
$T \in {\tt T}^{\rm ur}_n$, let 
\[
E\(T\) := \bigotimes_{v \in
\vert(T)}E\(\leg(v)\),
\] 
where $\leg(v)$ is the set of all edges of
$T$ adjacent to the vertex $v$. 

A {\em cyclic operad} is then a cyclic 
collection $\calC = \collc {\calC}$
together with a `coherent' system of `contractions'
\begin{equation}
\label{nigra_sum}
\alpha_T : \calC\(T\) \to \calC\(n\),\ T \in {\tt T}^{\rm ur}_n,\
n\geq 1,
\end{equation}
see~\cite[Definition~2.1]{getzler-kapranov:CPLNGT95}

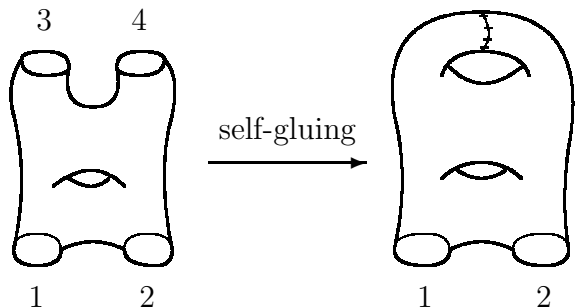
\begin{figure}
\begin{center}
{
\unitlength=0.300000pt
\begin{picture}(508.63,340.00)(0.00,0.00)

\put(-100,0){
\put(163.33,330.00){\makebox(0.00,0.00){$4$}}
\put(43.33,330.00){\makebox(0.00,0.00){$3$}}
\put(173.33,-20.00){\makebox(0.00,0.00){$2$}}
\put(33.33,-20.00){\makebox(0.00,0.00){$1$}}
\thicklines
\bezier{100}(103.33,220.00)(143.33,220.00)(133.33,270.00)
\bezier{100}(73.33,270.00)(63.33,220.00)(103.33,220.00)
\bezier{100}(43.33,290.00)(13.33,290.00)(13.33,280.00)
\bezier{100}(73.33,270.00)(73.33,290.00)(43.33,290.00)
\bezier{100}(43.33,260.00)(73.33,260.00)(73.33,270.00)
\bezier{100}(13.33,280.00)(13.33,260.00)(43.33,260.00)
\bezier{100}(163.33,290.00)(193.33,290.00)(193.33,280.00)
\bezier{100}(133.33,270.00)(133.33,290.00)(163.33,290.00)
\bezier{100}(163.33,260.00)(133.33,260.00)(133.33,270.00)
\bezier{100}(193.33,280.00)(193.33,260.00)(163.33,260.00)
\bezier{100}(203.33,210.00)(213.33,260.00)(193.33,280.00)
\bezier{100}(3.33,210.00)(-6.67,250.00)(13.33,280.00)
\bezier{100}(73.33,130.00)(103.33,110.00)(123.33,130.00)
\bezier{100}(53.33,120.00)(103.33,160.00)(143.33,120.00)
\bezier{100}(63.33,40.00)(103.33,60.00)(143.33,40.00)
\bezier{150}(203.33,210.00)(183.33,140.00)(203.33,40.00)
\bezier{150}(3.33,210.00)(23.33,130.00)(3.33,40.00)

\thinlines
\bezier{30}(63.33,40.00)(63.33,60.00)(33.33,60.00)
\bezier{30}(33.33,60.00)(3.33,60.00)(3.33,40.00)
\bezier{30}(143.33,40.00)(143.33,60.00)(173.33,60.00)
\bezier{30}(173.33,60.00)(203.33,60.00)(203.33,40.00)

\thicklines
\bezier{100}(33.33,20.00)(63.33,20.00)(63.33,40.00)
\bezier{100}(3.33,40.00)(3.33,20.00)(33.33,20.00)
\bezier{100}(143.33,40.00)(143.33,20.00)(183.33,20.00)
\bezier{100}(183.33,20.00)(203.33,20.00)(203.33,40.00)
}

\thicklines
\put(150,150){\vector(1,0){200}}
\put(250,190){\makebox(0,0){self-gluing}}

\put(100,0){
\put(473.33,-20.00){\makebox(0.00,0.00){$2$}}
\put(323.33,-20.00){\makebox(0.00,0.00){$1$}}
\thicklines
\bezier{100}(353.33,270.00)(393.33,230.00)(443.33,270.00)
\bezier{100}(363.33,140.00)(393.33,120.00)(423.33,140.00)
\bezier{100}(393.33,290.00)(443.33,290.00)(453.33,260.00)
\bezier{100}(343.33,260.00)(353.33,290.00)(393.33,290.00)
\bezier{100}(393.33,150.00)(423.33,150.00)(443.33,130.00)
\bezier{100}(343.33,130.00)(363.33,150.00)(393.33,150.00)
\bezier{150}(503.33,40.00)(493.33,100.00)(503.33,170.00)
\bezier{150}(393.33,340.00)(533.33,340.00)(503.33,170.00)
\bezier{150}(283.33,210.00)(263.33,340.00)(393.33,340.00)
\bezier{150}(283.33,210.00)(303.33,130.00)(283.33,40.00)
\bezier{150}(353.33,40.00)(393.33,60.00)(433.33,40.00)

\thinlines
\bezier{50}(393.33,340.00)(413.33,310.00)(393.33,290.00)
\put(398,320){\line(1,0){10}}
\put(392,335){\line(1,0){10}}
\put(397,305){\line(1,0){10}}
\put(393,295){\line(1,0){10}}

\thinlines
\bezier{30}(473.33,60.00)(503.33,60.00)(503.33,40.00)
\bezier{30}(433.33,40.00)(433.33,60.00)(473.33,60.00)
\bezier{30}(323.33,60.00)(353.33,60.00)(353.33,40.00)
\bezier{30}(283.33,40.00)(283.33,60.00)(323.33,60.00)

\thicklines
\bezier{100}(473.33,20.00)(433.33,20.00)(433.33,40.00)
\bezier{100}(503.33,40.00)(503.33,20.00)(473.33,20.00)
\bezier{100}(353.33,40.00)(353.33,20.00)(323.33,20.00)
\bezier{100}(283.33,40.00)(283.33,20.00)(323.33,20.00)
}
\end{picture}}

\end{center}
\caption{\label{lingua}
An example of `self-gluing.' The surface on the right has~$2$
punctures and genus~$2$. It is obtained from the surface on the left
with~$4$ punctures and genus~$1$ by sewing along the punctures marked
by~$3$ and~$4$.
} 
\end{figure}

Modular operads, anticipated in~\cite{behrend-manin:1995}, 
were introduced by
Getzler and Kapranov~\cite{getzler-kapranov:CompM98} 
for the study of moduli spaces of
Riemann surfaces of arbitrary genera.
Recall that a {\em modular collection\/} is a cyclic 
collection $E$ with a
second grading by the `genus' $g\geq 0$, $E = \collm
{E}$. 
A modular operad 
$\calA$ is then modular collection which posses, besides a 
cyclic operadic structure, also
operations $\calA \dz g{n+2} \to \calA \dz {g+1}n$. These operations are 
abstractions of the `self-gluing' which produces, from a
surface of genus $g$ with $(n+2)$ punctures, a new surface of genus
$g+1$ with $n$ punctures, as indicated in Figure~\ref{lingua}.

As cyclic operads are characterized by a system of
contractions~(\ref{nigra_sum}) indexed by unrooted trees, there is a
similar characterization of modular operads, but based
on labelled (or `modular') graphs rather than trees.

Following~\cite{behrend-manin:1995,kontsevich:94}, 
by a {\em graph\/} 
$\Gamma$ we mean a finite set $\Flag(\Gamma)$ (whose element are called
flags or half-edges) together with an involution 
$\sigma$ and a partition $\lambda$. 
The {\em vertices\/} $\Vert(\Gamma)$ of a graph 
$\Gamma$ are the blocks of the partition
$\lambda$.
The {\em edges\/}
$\Edg(\Gamma)$ are pairs of flags forming a two-cycle of $\sigma$
relative to the decomposition of a permutation into disjoint cycles. The
{\em legs\/} $\leg(\Gamma)$ are the fixed-points 
of $\sigma$.
We also denote by $\Leg(v)$ the flags belonging to the
block $v$ or, in common speech, half-edges adjacent to the vertex
$v$. 

Each graph $\Gamma$ has its {\em geometric
realization\/},
a finite one-dimensional cell complex
$|\Gamma|$, obtained by taking one copy of $[0,\frac 12]$ for each flag
and imposing the following equivalence relation: the points $0\in
[0,\frac 12]$ are identified for all flags in a block of the partition
$\lambda$, and the points $\frac 12 \in [0,\frac 12]$ are identified
for pairs of flags exchanged by the involution $\sigma$.
We will
usually make no distinction between a graph and its geometric realization. 

A  {\em modular\/} or 
{\em labelled\/} graph is a connected graph $\Gamma$ together with
a map $g:\Vert(\Gamma)\to \{0,1,2,\ldots\}$.
The {\em genus\/}
$g(\Gamma)$ of a modular graph $\Gamma$ is the number
\[
g(\Gamma) :=  \dim H_1(|\Gamma|)+\sum_{v\in \Vert(\Gamma)} g(v).
\]

Let $\Gammacat gS$ be the category whose objects are 
pairs $(|\Gamma|,\rho)$
consisting of a modular graph $\Gamma$
of genus $g$ and an isomorphism $\rho : \Leg(\Gamma) \to S$ labeling
the legs of $\Gamma$ by elements of a finite set $S$.
As usual, we write $\Gammacat gn := \Gammacat g{\{\rada 0n\}}$.

For a modular collection $\calA = \collm{\calA}$  and a 
modular graph $\Gamma$, let ${{\calA}}\(\Gamma\)$
be the tensor product
\begin{equation} 
\label{Cosmo}
{{\calA}}\(\Gamma\): = \bigotimes_{v\in\Vert(\Gamma)} 
{{\calA}}\(g(v),\Leg(v)\).
\end{equation}
A modular operad structure on $\calA$ is then given by a coherent system of
contractions~\cite[2.10]{getzler-kapranov:CompM98}
\[
\alpha_\Gamma : \calA\(\Gamma\) \to \calA \dz gS,
\mbox { for any $\Gamma \in \Gammacat gS$, $g\geq 0$ and a finite set $S$.}
\]

\begin{example}
\label{Kacirek_Donald}
{\rm\
Let $V = (V,B)$ be a differential graded 
vector space with a graded symmetric
inner product $B : V\ot V \to \bk$. 
Let us define, for each  $g \geq 0$ and a finite set $S$, 
\[
\End_V\(g,S\) := \otexp VS{}
\mbox { (the tensor product of copies of $V$ indexed by $S$).}
\]
It follows from definition that, 
for any graph $\Gamma\in \Gammacat gS$, 
$\End_V \(\Gamma\) = \otexp V{\Flag(\Gamma)}$. 

Let $\otexp B{\Edg(\Gamma)} : \otexp V{\Flag(\Gamma)} \to \otexp
V{\Leg(\Gamma)}$ be the multilinear form which contracts the factors
of $\otexp V{\Flag(\Gamma)}$ corresponding to the flags which are
paired up as edges of $\Gamma$. Then we define 
$\alpha_\Gamma: \End_V \dz g\Gamma \to \End_V \dz gS$ to be
the map
\begin{equation}
\label{Ja_jako}
\alpha_\Gamma :
\End_V\(\Gamma\) \cong \otexp V{\Flag(\Gamma)} 
\ \stackelra{\otexp B{\Edg(\Gamma)}}\  
\otexp
V{\Leg(\Gamma)} \cong \otexp VS{} = \End_V \dz gS.
\end{equation}
It is easy to show that the contractions $\{\alpha_\Gamma|\
\Gamma \in \Gammacat gS\}$
define on $\End_V$ the structure of a modular operad. 
}
\end{example}

We would like to modify Example~\ref{Kacirek_Donald} to the situation 
when the degree of the form $B$ is $+3$, as in the definition of a 
loop homotopy Lie algebra. Formula~(\ref{Ja_jako}) does not work,
among other things also because $\alpha_\Gamma$ will 
not be of degree zero.

For this modification 
we need to introduce `twisted' modular operads.
If  $X$ is a finite set with  ${\rm card}(X) = s$, 
let $\Det(X) := \ext^{s}((\desusp \bfk)^{\oplus X})$, the top dimensional
piece of the $s$-fold exterior power of the direct sum of the copies of
$\desusp \bfk$ indexed by elements of $X$. Clearly $\Det(X)$ is
an one-dimensional vector space concentrated in degree $-s$. 
The {\em determinant of a graph\/} $\Gamma \in \Gammacat gS$
is defined by $\Det(\Gamma) : =
\Det(\Edg(\Gamma))$. 

A {\em twisted\/} modular operad~(\cite[p.~293]{behrend-manin:1995}, 
also called a
$\frakK$-modular operad in~\cite{getzler-kapranov:CPLNGT95}) 
is then a modular 
collection $\calA$
together with a coherent system of contractions
\[
{\tilde \alpha}_\Gamma : \calA\(\Gamma\) \ot \Det(\Gamma)
\to \calA \dz gS,
\mbox { for any $\Gamma \in \Gammacat gS$, $g\geq 0$ and a finite set $S$.}
\]

\begin{example}
\label{Motylek}
{\rm
Let $W = (W,H)$ be a graded vector space with a 
nondegenerate degree~$-1$ symmetric 
bilinear form $H$. 
Define the modular collection
${\widetilde {\End}}_W$ by
\[
{\widetilde {\End}}_W\(g,S\) := \otexp WS,
\]
for $g\geq 0$ and a finite set $S$.
For $\Gamma \in \Gammacat gS$, the twisted modular contraction
\[
{\tilde{\alpha}}_\Gamma:
{\widetilde {\End}}_W\(\Gamma\) \ot {\Det }(\Gamma) 
\to {\widetilde {\End}}_W\(g,S\)
\]
is defined as follows. Let us choose 
labels ${s_e,t_e}$ 
such that $e = \{s_e,t_e\}$ for each edge $e \in \Edg(\Gamma)$
and define ${\tilde \alpha}_\Gamma$ to be
the composition:
\begin{eqnarray}
\nonumber  
\hskip 3mm
\hskip 5mm {\widetilde {\End}}_W\(\Gamma\) \ot {\Det }(\Gamma)
\hskip -2mm &\cong& \hskip -2mm \otexp W{\Flag(\Gamma)} 
\ot \Det(\Gamma)
\cong
\otexp WS
\ot \hskip -4mm
\bigotimes_{e\in \Edg(\Gamma)} \hskip -2mm
\lbigbrace
\otexp W{\{s_e,t_e\}} \ot \Span(\desusp e)
\rbigbrace
\\
\nonumber 
\hskip -2mm &\cong& \hskip -2mm
\otexp WS
\ot
\bigotimes_{e\in \Edg(\Gamma)} 
\lbigbrace 
W_{s_e} \ot W_{t_e}
\ot
\Span(\desusp e)
\rbigbrace
\\
&&
\nonumber 
\stackelra{\id \ot\bigotimes_{e}H_e}\
\otexp W{S} \ot \otexp \bk{\Edg(\Gamma)} 
\cong {\widetilde {\End}}_W \dz gS,
\end{eqnarray}
where $H_e$ is the map that sends $u \ot v \ot \hskip -1.5mm \desusp
\hskip -1mm e 
\in W_{s_e} \ot W_{t_e}
\ot
\Span(\desusp \hskip -1mm e)$ to $H(u,v) \in \bfk$.
The symmetry 
of $H$ assures that the
the definition of ${\tilde\alpha}_\Gamma$ does not
depend on the choice of labels.
The system $\{{\tilde\alpha}_\Gamma|\
\Gamma \in \Gammacat gS\}$ 
induces on ${\widetilde {\End}}_W$ the structure of a twisted
modular operad.
}
\end{example}

If $V = (V,B)$ is a graded vector space 
with a nondegenerate degree $+3$ bilinear symmetric form $B$, then $W
= (W,H)$ with $W := \suspit 2 V$ and the form $H$ defined by
$H(u,v) := B(\desuspit 2 u, \desuspit 2 v)$, $u,v \in W$, form the data
as in Example~\ref{Motylek}, so we may consider
the twisted modular operad ${\widetilde {\End}}_{\suspit 2 V}$.

Another example of a twisted modular operad is provided by the {\em Feynman
transform\/} of a modular operad.  
Recall~\cite[4.2]{getzler-kapranov:CompM98} that the {\em free
twisted modular operad\/} $\twModtriple(E)$
on a modular collection $E$ is given by
\[
\twModtriple(E)\(g,n\) :=
\colim{\Gamma \in \isogamma{}gn} E\(\Gamma\) \ot \Det(\Gamma),
\]
where $\isogamma{}gn$ is the full subcategory of isomorphisms in
$\Gammacat gn$.
The twisted modular operad structure is induced by the `grafting' of
underlying graphs.

If $\calA$ is a modular operad, 
then $\twModtriple(\calA)\(g,n\)$ carries
a natural differential $\pa_\fey$~\cite[Theorem~4.4]{behrend-manin:1995}. 
The twisted  differential modular operad
$\Fey(\calA) := (\twModtriple(\calA),\pa_{\fey})$ is called the
Feynman transform of the modular operad $\calA$.   

Let us consider the `forgetfull' 
functor $\Box : \MOp \to \COp$ from the category
of modular operads to the category of cyclic operads given
by $\Box(\calA)\(S\) := \calA\(0,S\)$, for any
finite set $S$. It is not difficult to 
show~\cite{markl-shnider-stasheff:book}
that this functor has a left
adjoint $\Mod : \COp \to \MOp$. 

\begin{definition}
\label{Andulka}
The modular operad $\Mod(\P)$ is called 
the {\em modular operadic completion\/} of the cyclic operad
$\P$.
\end{definition}

An easy calculation shows that
\begin{equation}
\label{HaFiK}
\Mod(\Com) \dz gn \cong \bk, \mbox{ for each $g\geq 0$, $n\geq 1$,}
\end{equation}
with the trivial action of the
symmetric group $\Sigma_{n+1}$. 

The key role in our characterization is played by the Feynman
transform $\Fey(\ModCom)$ of the modular completion of the operad
$\Com$. It follows from~(\ref{HaFiK}) that, as a nondifferential
operad, $\Fey(\ModCom)$ is the free twisted modular operad on the
generators $\omega^g_n$,
\begin{equation}
\label{Jezecek}
\twModtriple(\Mod(\Com)) \cong
\twModtriple(
              \{
                 \omega^g_n;\ n\geq 1, g\geq 0
              \}
             ),
\end{equation}
where $\omega^g_n$ corresponds 
to the dual of $1 \in \bfk \cong \Mod(\Com) \(g,n\)$. 
The central result of this section reads as follows.

\begin{theorem}
\label{Jituska}
There exists a natural one-to-one correspondence between twisted modular
$\Fey(\ModCom)$-algebra structures on 
$(\susp^2V,B(\desusp^2-,\desusp^2-)$, i.e.~morphisms
\begin{equation}
\label{tezko}
a : \lbigbrace
          \Fey(\ModCom),\pa_\fey
    \rbigbrace 
                           \to 
    \left(
         {\widetilde {\End}}_{\uparrow^2 V},\pa=0
    \right)
\end{equation}
of differential 
twisted modular operads, and loop homotopy algebra structures on
$(V,B)$ in the sense of Definition~\ref{e}.
\end{theorem}

\noindent
{\bf Sketch of proof.}
Description~(\ref{Jezecek}) shows  that a map $a$ of~(\ref{tezko}) 
is determined by its values $\xi^g_n : =
a(\omega^g_n) \in {\widetilde {\End}}_{\uparrow^2 V}\(g,n\)$ 
on the generators. Moreover, the map $a$ ought to commute with
the differentials, so the equation
\begin{equation}
\label{hvezdicka}
a(\pa_\Fey(\omega^g_n)) = 0
\end{equation}
must be satisfied, for each $g\geq 0$ and $n \geq 1$. 
Observe that $\xi^g_n \in {\widetilde {\End}}_{\uparrow^2 V}
\(g,n\)$ can be
interpreted as a degree $-2(n+1)$-element of the graded vector space
$\otexp V{n+1}$.   
Let us introduce  a map 
$\Xi : \otexp V{n+1} \to \Hom{\otexp Vn}V$ by
\begin{eqnarray}
\label{Vrana}
\lefteqn{
\Xi(\Orada x0n)(\Rada v1n) :=
}
\\
\nonumber&& 
:=
\znamenko{nx_0 + (n-1)x_1 + \cdots +x_{n-1}}
x_0 B(x_1,v_1)B(x_2,v_2) \cdots B(x_n,v_n),
\end{eqnarray}
for $\Orada x0n \in \otexp V{n+1}$ and $\Rada v1n \in V$. The map $\Xi$
is clearly a degree $3n$ isomorphism of $\otexp V{n+1}$ and
$\Hom{\otexp Vn}V$.
Finally, let $l^g_n : \otexp Vn \to V$ be a homogeneous degree $n-2$ 
map given by
\[
l^g_n(\Rada v1n) := \znamenko{\frac{n(n+1)}2 + n(\Prada v1n)} \hskip 1mm 
\Xi(\omega^g_n)(\Rada v1n), 
\mbox { for $\Rada v1n \in V$}.
\]

A long but straightforward calculation shows that $l^g_n$ 
are antisymmetric
operations satisfying~(\ref{acx}) 
and that~(\ref{hvezdicka}) translates to the main
identity~(\ref{6}).

On the other hand, all steps above can clearly be reversed,
thus a loop homotopy Lie algebra structure induces a map~(\ref{tezko}). 
\qed

\begin{remark}
{\rm
Observe that Theorem~\ref{Jituska} is formulated in such a way that
the differential $\pa$ on $V$ is a part of the structure, namely $\pa
:= a(\omega^0_1)$. 
\/}
\end{remark}

\section{Possible generalizations (open strings)}
\label{generalizations}

Let $\P$ be an operad. It is now well-understood what a `strongly homotopy
$\P$-algebra' is. In case when $\P$ is Koszul, 
it is an algebra over the cobar
construction on the quadratic dual $\P^!$ of 
$\P$~\cite[Definition 4.2.14]{ginzburg-kapranov:DMJ94}. 

An alternative characterization is that a homotopy 
$\P$-algebra is a square zero differential on
the cofree connected $\P^!$-coalgebra. The equivalence of these 
two characterizations
follows for example
from~\cite[Proposition~4.2.15]{ginzburg-kapranov:DMJ94}.  

The quadratic dual of the operad $\Lie$ for Lie algebras is $\Com$,
the operad for commutative associative algebras, and
the above characterization give
Proposition~\ref{wv} resp.~%
Proposition~\ref{vw}.
Another example is $\P=\Ass$, the operad for associative algebras. 
It is quadratic self-dual, $\P^! =
\Ass$, and the corresponding strongly homotopy algebras 
are called  strongly homotopy associative or 
$A_\infty$-%
algebras~\cite{stasheff:TAMS63,markl:JPAA92}.

Let us look for possible generalizations to the loop case.
If $P$ is a {\em cyclic\/} operad
(recall that both $\Lie$ and $\Ass$ are cyclic),
the quadratic dual $\P^!$ is again 
cyclic~\cite{getzler-kapranov:CPLNGT95}, so it makes sense to consider the
modular completion $\Mod(\P^!)$ (Definition~\ref{Andulka}). 
We suggest the following definition.

\begin{definition}
Let $\P$ be a Koszul cyclic operad. A {\em loop homotopy
$\P$-algebra\/} 
is then a modular algebra over the twisted differential modular operad
$\Fey({\Mod(\P^!)})$.
\end{definition} 

For $\P=\Lie$ we get Theorem~\ref{Jituska}.
It would  be interesting to write out explicitly axioms of 
{\em loop homotopy associative algebras\/}, because these structures 
should play an
important r\^ole in the higher-genera open string field theory,
as suggested by~\cite{stasheff3}. While in the 
Lie case we had, for each $n$ and $g$, only one
antisymmetric operation $l^g_n: V^{\ot n}\to V$, 
in the loop homotopy associative case we expect to
have
\[
\frac{(n+1)!}{2^g \cdot g!\cdot (n+1-2g)!}
\]
operations $V^{\ot n}\to V$, due to the dimension of
$\Mod(\Ass)\(g,n\)$.

A seemingly easier approach would be the one 
based on coderivations. We would
like to say that a loop homotopy $\P$-algebra is an order 2
coderivation of the cofree connected $\P^!$-coalgebra, having
properties analogous to~(\ref{711}). 
This works nicely for $\P = \Lie$, because we know what
is a higher order coderivation of a cocommutative coalgebra. But we
are not sure whether there exist a reasonable concept of
higher-order coderivations without the cocommutativity, though the
paper~\cite{alfaro-damgaard} seems to suggest this.


\catcode`\@=11

\noindent
Mathematical institute of the Academy, 
\v Zitn\'a 25, 
115 67 Prague 1, 
The Czech Republic\hfill\break
e-mail: {\tt markl@math.cas.cz}

\end{document}